\documentclass[a4paper, 10pt, twocolumn]{article}

\usepackage[T1]{fontenc}
\usepackage{amssymb}
\usepackage{amsfonts}
\usepackage{amsmath}

\usepackage{graphicx}

\newcommand{\om}{\omega}

\title{A Simple Denoising Technique}
\author{    F. Douarche, L. Buisson, S. Ciliberto, A. Petrosyan \\
            Laboratoire de Physique de l'ENS Lyon --- CNRS UMR 5672 \\
            46, All\'ee d'Italie --- 69364 Lyon Cedex 07, France}

\begin{document}

\maketitle

The measurements of very low level signals at low frequency is a
very difficult problem, because environmental noise increases in
this frequency domain and it is very difficult to filter it
efficiently. In order to counteract these major problems, we
propose a simple and generic denoising technique, which mixes
several features of traditional feedback techniques and those of
noise estimators. As an example of application, large band
measurements of the thermal fluctuations of a mechanical
oscillator are presented. These measurements show that the proposed
denoising technique is easy to implement and gives good results.

\section{Introduction} \label{sec1}

The measurements of very low level signals at very low frequency
(VLF) is a very difficult problem, because environmental and
electric\,/\,magnetic noises often increase in this frequency domain.
Furthermore, it is well known that the difficulties of
isolating an experimental setup from these unwanted noise
sources increase by reducing the measuring frequency. Typical
examples are the screening of low frequency magnetic fields or
the isolation of a measurement from the unwanted environmental
vibrations. Many techniques have been proposed and accurately
applied to reduce the effects of the unwanted noise sources.

The simplest techniques are of course the passive ones. Let us
consider in some details the problem of vibration isolation (the
magnetic field screening presents similar problems).

In a typical laboratory environment, vibrations transmitted through the
floor normally have a frequency spectrum from dc to a few hundred Hz.
The reduction of these vibrations is usually obtained by installing
the experiment on floating tables which have horizontal and vertical
resonance frequencies around 1 Hz. Thus, noise reduction is obtained
only for frequencies larger than the natural resonant frequencies of
the table. This is, of course, an excellent method for high frequency
measurements, but at frequencies close to and smaller than 1 Hz this
method becomes useless. Exactly at resonance, noise is even enhanced.
To overcome these problems that appear at VLF, feedback techniques have
been used. These techniques require detectors which measure the noise
signals and actuators which reduce the acceleration of the table plate.
Similar techniques are of course used for screening VLF magnetic fields.
Indeed, high frequency magnetic fields are screened by Faraday cages and
the VLF components are subtracted by a feedback technique \cite{lindgren}.

Feedback techniques are widely used, but they are limited by the
noise of the detectors and of the actuators. Their calibration is
often very complex and requires very tedious operations in order
to work properly. Furthermore, they can be only applied in all of
the cases where the environmental noise can be accurately measured.
When this is not possible and\,/\,or one is not interested in stabilizing
a system on a given working point, but only in reducing the noise
on a given signal, other techniques may be used. From the signal
analysis point of view, most of the techniques that have been
proposed and accurately applied to reduce the effects of the
unwanted noise sources rely upon the knowledge of the response
function of the system under study, and a guess about the noise,
which is often supposed to be a random variable belonging to some
class of signals, e.g. ergodic and second order stationary signals
\cite{papoulis, max, lifermann, kalman, kalmanb}. However, this guess is often
limited.

The simple denoising technique proposed in this paper actually
combines several aspects of the two above mentioned techniques.
Indeed, in many cases one has access to the environmental noise,
but one does not need to stabilize a setup on a given working
point, but only to reduce the effect of the noise on a given
signal. Therefore, rather than carrying out a somewhat sophisticated
and expensive feedback system, which could even fail to solve the
noise problem at VLF, we developped a simple denoising technique whose
principle precisely lies in measuring the residual noise when passive
``screening'' devices are already used. The general principles and
the limits of the new technique are described in Sec.\,\ref{sec2}.

This denoising technique was motivated by the study of the violation
of the fluctuation dissipation theorem (FDT) in out of equilibrium
systems as aging glasses. This is a subject of current interest
which begins to be widely studied in many different systems
\cite{israeloff, be3, herisson, be5, lionel}.
Therefore, in Sec.\,\ref{sec3} we propose an application of this
new technique to the measurement of VLF mechanical thermal
fluctuations. The experimental results, presented in Sec.\,\ref{sec4},
show the quality of the noise reduction. Finally, in Sec.\,\ref{sec5}
we discuss some other possible applications and we conclude.

\section{A simple denoising technique} \label{sec2}

Suppose one has to measure a signal on which an external noise is
superimposed. Let us call this signal the true signal $x_{\mathrm{true}}(t)$
and the external noise $x_{\mathrm{env}}(t)$, so that the measured signal
can be written in the additive manner
\begin{equation}
    x(t) = x_{\mathrm{true}}(t) + x_{\mathrm{env}}(t). \label{eq00}
\end{equation}

\noindent The only assumptions needed are that $x_{\mathrm{true}}(t)$
and $x_{\mathrm{env}}(t)$ are uncorrelated and both stationary processes (as
we will see, the hypothesis of stationarity can be weakened), and can be written in
an additive manner like in Eq.\,\ref{eq00}. In addition, we assume that one can
directly measure $x$, whereas $x_{\mathrm{env}}$ is measured with a detector
whose output signal is $x_{\mathrm{det}}(t)$. If the noise is small, in the
limit of linear response theory, it can be stated that $x_{\mathrm{det}}$
is linearly related to $x_{\mathrm{env}}$, which means that there exists
a hypothetical response function $\hat{R}_{\mathrm{d}}^{-1}(\om)$
such that\begin{equation}
    \hat{x}_\mathrm{det}(\om) = \hat{R}_{\mathrm{d}}^{-1}(\om)
    \hat{x}_{\mathrm{env}}(\om),
\end{equation}

\noindent where $\hat{f}(\om) = \int_{\mathbb{R}} f(t) e^{-i \om t} dt$ is
the Fourier transform of $f(t)$. Notice that no hypothesis is done on
$\hat{R}_{\mathrm{d}}$, which is in principle unknown.

The Fourier transform of the true signal thus reads
\begin{equation}
    \hat{x}_{\mathrm{true}} = \hat{x} - \hat{R}_{\mathrm{d}} \,
    \hat{x}_{\mathrm{det}}. \label{eq0}
\end{equation}

\noindent Assuming that the true and external noise signals are uncorrelated,
that is $\langle x_{\mathrm{true}} \, x_{\mathrm{det}}^{*} \rangle = 0$,
one can compute the kernel $\hat{R}_{\mathrm{d}}$ as
\begin{equation}
    \hat{R}_{\mathrm{d}} = \frac{\langle \hat{x}
    \, \hat{x}_{\mathrm{det}}^{*} \rangle}{\langle {\vert
    \hat{x}_{\mathrm{det}} \vert}^2 \rangle}, \label{eq1}
\end{equation}

\noindent where $\langle \cdot \rangle$ stands for the ensemble average.
Thus Eqs.\,\ref{eq0} and \ref{eq1} allow us to compute the signal and its
spectrum:
\begin{equation}
    x_{\mathrm{true}}(t) = x(t) - \int_{\mathbb{R}} \hat{R}_{\mathrm{d}}
    (\om) \hat{x}_{\mathrm{det}} (\om) e^{i \om t} \, \frac{d \om}{2 \pi},
    \label{eq3}
\end{equation}

\noindent and
\begin{equation}
    \langle {\vert \hat{x}_{\mathrm{true}} \vert}^2 \rangle = \langle
    {\vert \hat{x} \vert}^2 \rangle - {\vert \hat{R}_{\mathrm{d}} \vert}^2 \langle
    {\vert \hat{x}_{\mathrm{det}}\vert}^2 \rangle. \label{eq2}
\end{equation}

\noindent Therefore, $x_{\mathrm{true}}$ can be computed from the simultaneous
measurements of  $x$ and $x_{\mathrm{env}}$.

We see that the hypothesis of stationarity is not really
necessary, because $\hat{x}$, $\hat{R}_{{d}}^{-1}$ and
$\hat{x}_{{det}}$ can be slowly varying functions of time, with a
characteristic time $\tau$. In such a case, if the ensemble
average is performed in a time ${T}$ such that ${T} \ll \tau$,
then Eqs.\,\ref{eq0}, \ref{eq1}, \ref{eq3} and \ref{eq2} can be
still applied on intervals of length ${T}$. This observation makes
this simple technique very powerful, because the response of the
system to the environmental noise can change as a function of time
and of the external noise source. Thus, $\hat{R}_{\mathrm{d}}$ is
a dynamical variable which can be computed in each time interval
of length $ {T}$, and which allows to retrieve the true signal.

However, the signal $x_{\mathrm{rec}}$ reconstructed using
Eqs.\,\ref{eq0}, \ref{eq1}, and \ref{eq3} will differ from
$x_{\mathrm{true}}$ because of experimental errors. One source of
error is the noise of the detectors and of the amplifiers, which
introduces an extra additive noise term $\eta(t)$ in
Eq.\,\ref{eq00}, which is uncorrelated with $x_{\mathrm{true}}$
and $x_{\mathrm{det}}$, thus $x_{\mathrm{rec}}(t) =
x_{\mathrm{true}}(t) + \eta(t)$. However, $\eta$ can be done very
small and it does not constitute the main source of error. The
main one is the limited number $N$ of ensemble averages that can
be done in the time $ {T}$. This is very important because if
$\hat{R}_{\mathrm{d}}$ is a slowly varying function of $t$, then
one has to impose $ {T} \ll \tau$ in order to retrieve the true
signal. Finally, it has to be pointed out that the advantage of
the technique is when the amplitudes of $x_{\mathrm{true}}$ and
${x}_{\mathrm{det}}$ are comparable, that is, when the signal to
environmental noise ratio is either smaller than or equal to one.

The accuracy of the technique has been checked on several artificial
signals. We have chosen for $x_{\mathrm{true}}$ and ${x}_{\mathrm{det}}$
either random or periodic signals. The random signals may be either
colored or white noise with Gaussian or uniform distribution.
To estimate the error of the reconstruction, we first consider the
difference $\Delta$ between the reconstructed $x_{\mathrm{rec}}$
and $x_{\mathrm{true}}$, that is
$\Delta(t) = x_{\mathrm{rec}}(t) - x_{\mathrm{true}}(t)$.
We then compute the ratio $R_{\mathrm{rec}}$ between the rms of $\Delta$
and that of $x_{\mathrm{true}}$, which is a good indicator of the
error of the reconstructed signal. As expected we find that
$R_{\mathrm{rec}} \sim 1/ \sqrt{N}$ for large $N$.

\begin{figure}
    \begin{center}
    \includegraphics[width=8cm, angle=0]{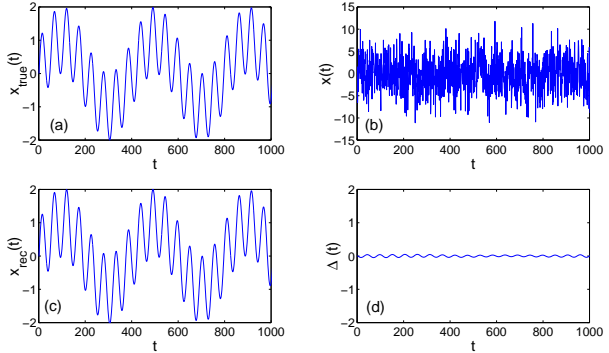}
    \caption{Artificial signal. (a) True signal as a function of time.
    In this example $x_{\mathrm{true}}(t) = \sin{\frac{2 \pi}{51} t} \, +
    \sin{\frac{2 \pi}{400} t}$. (b) The signal $x$ as a function of time.
    The noise is a Gaussian white noise of variance 4. (c) The
    reconstructed signal as a function of time. (d) Residual error
    $\Delta$ of the reconstruction as a function of time.}
    \label{fig1}
    \end{center}
\end{figure}

An example of the reconstruction is given in Fig.\,\ref{fig1}. We see that
although the signal is completely erased by the noise (cf Fig.\,1b)
the reconstruction is quite good. It is obvious that this is an
extremely simple example, but as we will see the technique becomes
very interesting when $\hat{R}_{\mathrm{d}}$ is a slowly varying
function of $t$.

To conclude this section, it should be stressed that, from the signal
analysis point of view, we derived a method in a way similar to the Wiener
filtering, which aim at separating (in an optimal sense, see \cite{papoulis,
max, nr}) two random signals $x_{\mathrm{true}}$ an $x_{\mathrm{det}}$, which are
supposed to be ergodic second order stationary and uncorrelated random signals,
and can be written in an additive manner like in Eq.\,\ref{eq00}. Then, we
extended this denoising technique to nonstationary signals in a simple and
original manner. In a more general study, nonstationary signals could be
addressed to the Kalman filtering (also referred as to the Kalman-Bucy
filtering), which can be considered as the extension of the Wiener filtering
to the case of nonstationary signals  \cite{lifermann, kalman, kalmanb}.

\section{Application to thermal fluctuations measurements} \label{sec3}

In this section, we describe a useful application of this technique
to the measurement of thermal fluctuations of a mechanical oscillator,
whose damping is given by the viscoelasticity of an aging polymer
glass. This is an important experimental measurement which is
extremely useful in the study of the violation of the FDT in out of
equilibrium systems, specifically in aging glasses. This violation
is a subject of current interest which has been studied mainly
theoretically \cite{cugliandolo, sollich}.
However, there are not clear experimental tests of
these theoretical predictions, which have to be checked on real
systems by studying the VLF spectra of mechanical thermal
fluctuations. Thus, the main purpose of our study is to have a
reliable measurement of this VLF spectra in an aging polymer.

To study this spectrum, we have chosen to measure the thermally
excited vibrations of a plate made of an aging polymer such as
Polycarbonate. The physical object of our interest is a small
plate with one end clamped and the other free, i.e. a cantilever. The
plate is of length $l$, width $a$, thickness $b$, mass
$m_{\mathrm{Polyc}}$. On the free end of the cantilever a small
golden mirror of mass $m_{\mathrm{mirror}}$ is glued. As described
in the next section, this mirror is used to detect the amplitude
$x_c$ of the transverse vibrations of the cantilever free end. The
motion of the cantilever free end can be assimilated to that of a
driven harmonic oscillator, which is damped only by the
viscoelasticity of the polymer. Therefore, the Fourier-transformed
equation of motion of the cantilever free end reads
\begin{equation}
    [- m \om^2 + K(\om) ] \hat{x}_c = \hat{F}_{\mathrm{ext}},
\label{HO}
\end{equation}

\noindent where $\hat{x}_c$ is the Fourier transform of $x_c$, $m$
is the total effective mass of the plate plus the mirror,
$K = K' + i K''$ is the complex elastic stiffness of the plate free
end, and $\hat{F}_{\mathrm{ext}}$ is the Fourier transform of the external
driving force. The complex $K(\om)$ takes into account the
viscoelastic nature of the cantilever. From the theory of
elasticity \cite{bib1} one obtains that, for VLF, excellent
approximations for $m$ and $K$ are:
\begin{eqnarray}
    m & = & \frac{3}{(3.52)^2} \, m_{\mathrm{Polyc}} + m_{\mathrm{mirror}}, \\
    \textrm{and } K & = & {E a b^3 \over 4 l^3},
\end{eqnarray}

\noindent where $E = E' + i E''$ is the plate Young modulus.
Notice that if $m_{\mathrm{mirror}} = 0$, then one recovers
the smallest resonant frequency of the cantilever \cite{bib1}.
For Polycarbonate at room temperature, $E$ is such that
$E' = 2.2 \times 10^9 \textrm{ Pa}$ and $E'' = 2 \times 10^7 \textrm{ Pa}$,
and its frequency dependence may be neglected in the range of frequency
of our interest, that is from 0.1 to 100 Hz \cite{bib3}. Thus we neglect the
frequency dependence of $K$ in this specific example.

When $F_{\mathrm{ext}} = 0$, the amplitude of the thermal vibrations of
the plate free end $x_T$ is linked to its response function
$\chi$ via the FDT \cite{bib2}:
\begin{equation}
    \langle \vert{ \hat{x}_T \vert}^2 \rangle = \frac{2 k_B T}{\om} \, \mathrm{Im} \, \hat{\chi}, \label{FDT}
\end{equation}

\noindent where $\langle \vert{ \hat{x}_T \vert}^2 \rangle $ is
the thermal fluctuation spectral density of $x_c$, $k_B$ the
Boltzmann constant and $T$ the temperature. From Eq.\,\ref{HO} one
obtains that the response function of the harmonic oscillator is
\begin{equation}
    \hat{\chi} = \frac{\hat{x}_c}{\hat{F}_{\mathrm{ext}}}
    = { 1 \over m \lbrack {\om_0}^2 - \om^2 - i \, (\mathrm{sign}\, \om) \, \gamma {\om_0}^2 \rbrack}, \label{eq4}
\end{equation}

\noindent where ${\om_0}^2 = K' / m$ and $\gamma = K'' / K'$.

Inserting Eq.\,\ref{eq4} into Eq.\,\ref{FDT}, one can compute the
thermal fluctuation spectral density of the Polycarbonate
cantilever for positive frequencies:

\begin{equation}
    \langle \vert{ \hat{x}_T \vert}^2 \rangle = \frac{2 k_B T}{\om}
    \frac{\gamma {\om_0}^2}{m \lbrack ({\om_0}^2 - {\om}^2)^2 + (\gamma {\om_0}^2)^2 \rbrack}.
    \label{eq5}
\end{equation}

\noindent Notice that $\langle {\vert{ \hat{x}_T }\vert}^2 \rangle
\sim \om^{-1}$ for $\om \ll \om_0$, because the viscoelastic
damping $K''$ is constant in our frequency range. In the case of a
viscous damping (for example, a cantilever immersed in a viscous
fluid) $K'' = \alpha \, \om$, where $\alpha$ is proportional to
the fluid viscosity and to a geometry dependent factor. Then the
spectrum of the thermal fluctuations of the cantilever free end,
in the case of viscous damping, is
\begin{equation}
    \langle {\vert \hat{x}_T \vert}^2 \rangle = \frac{2 k_B T
    \, \alpha}{m^2 \lbrack (\om_0^2-\om^2)^2 + (\frac{\alpha}{m} \om)^2
    \rbrack }, \label{eq5b}
\end{equation}

\noindent which is constant for $\om \ll \om_0$. Therefore the fluctuation
spectrum shape depends on $K''(\om)$. In the case of a viscoelastic damping
(see Eq.\,\ref{eq5}), the thermal noise increases for $\om \ll \om_0$, and
with a suitable choice of the parameters the VLF spectrum of an
aging polymer can be computed using this method.

However, the cantilever is also sensitive to the mechanical noise,
and the total displacement $x_c$ of the cantilever free end actually reads
$x_c = x_T + x_{\mathrm{acc}}$, where $x_{\mathrm{acc}}$ is the displacement induced
by the external mechanical noise. Thus, it is important to compute
the signal-to-noise ratio of our physical apparatus, which we define as
the ratio between the thermal fluctuations and the mechanical noise spectra.
To compute the latter, we consider that the support of the cantilever is
submitted to an external acceleration $a_{\mathrm{ext}}$, whose Fourier
transform is $\hat{a}_{\mathrm{ext}}$. We rewrite Eq.\,\ref{eq4}
with $\hat{F}_{\mathrm{ext}} = m \hat{a}_{\mathrm{ext}}$, which yields
\begin{equation}
    \hat{x}_{\mathrm{acc}} = \frac{\hat{a}_{\mathrm{ext}}}{{\om_0}^2
    - \om^2 - i\gamma \om_0^2}, \label{FFTac}
\end{equation}

\noindent where $\hat{x}_{\mathrm{acc}}$ is the Fourier transform
of $x_{\mathrm{acc}}$. Far from the resonance frequency, that
is for $\om \ll \om_0$, one has $\hat{x}_{\mathrm{acc}} \sim
\hat{a}_{\mathrm{ext}} / {\om_0}^2$, which finally yields ${\vert
\hat{x}_{\mathrm{acc}}\vert}^2 \sim {\vert \hat{a}_{\mathrm{ext}}
\vert}^2 / {\om_0}^4$, whereas the thermal fluctuation spectral
density of $x$ reads $\langle {\vert{ \hat{x}_T }\vert}^2 \rangle
\sim \frac{2 k_B T}{\om} \frac{\gamma}{m {\om_0}^2}$. Therefore, the
signal-to-noise ratio reads
\begin{equation}
    \frac{\langle{\vert{ \hat{x}_T }\vert}^2
    \rangle}{{\langle\vert{ \hat{x}_{\mathrm{acc}} }\vert}^2 \rangle} \sim
    \frac{2 k_B T}{\om} \frac{\gamma \om_0^2}{
     m {\langle \vert \hat{a}_{\mathrm{ext}} \vert}^2 \rangle},
    \label{SNR}
\end{equation}

\noindent which is proportional to
$\frac{\gamma E a b^3}{m^2 l^3 {{\langle \vert{ \hat{x}_{\mathrm{acc}} } \vert}^2 \rangle}}$,
for $\om \ll \om_0$. Notice that the signal-to-noise ratio of Eq.\,\ref{SNR}
increases if the set of parameters $\{ a, b, l, m \}$ is optimized to make
$\om_0$ as large as possible within the frequency range of interest, and
within the experimental constraints.

Let us estimate the amplitude of $\sqrt{ {\langle {\vert{ \hat{x}_T }\vert}^2 \rangle} }$ at
$\nu = \om / 2\pi = 1 \textrm{ Hz}$ for the following choice of the parameters:
$\gamma \simeq 10^{-2}$, $l \simeq 10 \textrm{ mm}$,
$a \simeq 1 \textrm{ mm}$, $b = 125\ \mu \mathrm{m}$ and
$m_{\mathrm{mirror}} \lesssim 10^{-3} \textrm{ g}$. We find
$\nu_0 \simeq 100 \textrm{ Hz}$ and
$\sqrt{ {\langle{ \vert{ \hat{x}_T (1 \textrm{ Hz})} \vert}^2 \rangle} } \simeq 10^{-11} \textrm{ m} / \sqrt{\mathrm{Hz}}$,
which is a very small signal. As a consequence, extremely small vibrations of the environment may greatly
perturb the measurement. Therefore, to increase the signal-to-noise ratio of the measurements,
one has to reduce the coupling of the cantilever to the environmental noise (acoustic and seismic)
using vibration isolation systems. This may be not enough in this specific case because of the the smallness
of the thermal fluctuations. Then we have applied the technique described in the previous section in order
to recover $x_T$ from the measurement of $x_c$. The experimental results are described in the next section.

\section{Experimental results} \label{sec4}

\begin{figure}
    \begin{center}
    \includegraphics[width=7cm, angle=0]{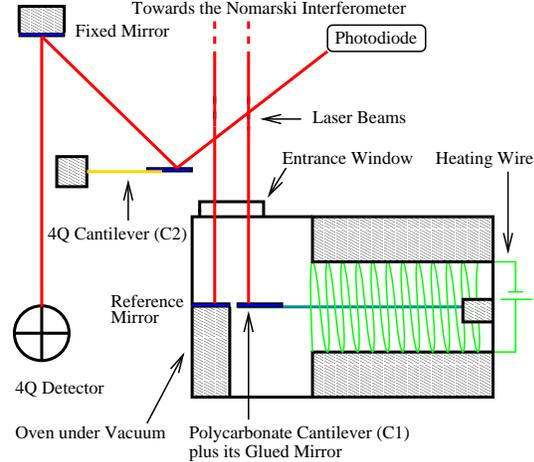}
    \caption{Experimental setup.} \label{fig2}
    \end{center}
\end{figure}

The measurement of $x_c$ is done using a Nomarski interferometer
(for detailed reviews, see \cite{bib4, bib5, bib6}) which uses the
mirror glued on the Polycarbonate cantilever in one of the two
optical paths. The interferometer noise is about
$5 \times 10^{-14} \textrm{ m} / \sqrt{\mathrm{Hz}}$, which is two orders
of magnitude smaller than the cantilever thermal fluctuations.
The cantilever is inside an oven under vacuum. A window allows
the laser beam to go inside (cf Fig.\,\ref{fig2}). The size of the
Polycarbonate cantilever are, $l \simeq 13.5 \textrm{ mm}$, $a \simeq 1 \textrm{
mm}$ and $b = 125\ \mu \mathrm{m}$, and the mirror mass is $m_{\mathrm{mirror}}
\lesssim 10^{-3} \textrm{ g}$ such that  $\nu_0 \simeq 100 \textrm{ Hz}$.

Much care has been taken in order to isolate as much as possible the
apparatus from the external mechanical and acoustic noise. The
Nomarski interferometer and the cantilever are mounted on a plate
which is suspended to a pendulum whose design has been inspired
by one of the isolating stages of the VIRGO superattenuator
\cite{bib7, bib8, bib9}. The whole ensemble is enclosed in a
cage, to avoid any acoustic coupling. The pendulum and the cage are
installed on air-suspended breadbord (Melles Griot Small Table Support System
07 OFA Series Active Isolation), which furnishes an extra
isolating stage. However, these two isolation stages are not yet
enough to have a large band measurement (0.1-100 Hz) of the
cantilever thermal fluctuations.

\begin{figure}
    \begin{center}
    \includegraphics[width=8cm, angle=0]{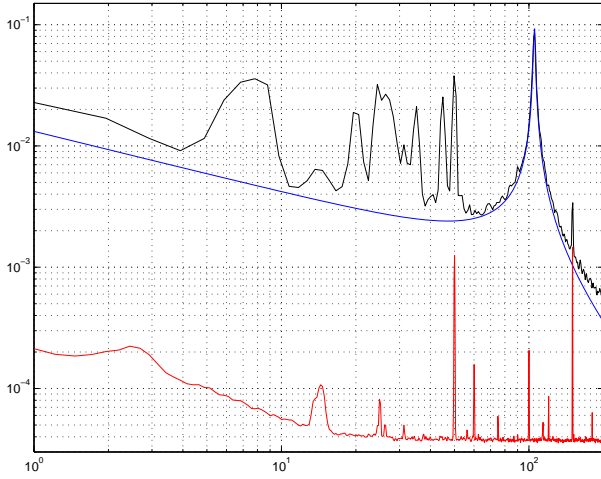} 
    \caption{Square root of the fluctuation spectral density without denoising
    ($\textrm{nm} / \sqrt{\mathrm{Hz}}$) vs frequency (Hz).}
    \end{center}
    \label{fig3}
\end{figure}

In Fig.\,\ref{fig3} we plot the square root of the cantilever's
fluctuation spectral density as a function of frequency for a
typical experiment at ambient temperature. The measure is compared
with the FDT prediction, obtained from Eq.\,\ref{FDT} (blue line), and
the interferometer noise (red line). The measure scales quite
well with the prediction. One can observe the cantilever resonance
and the $1 / \sqrt{\nu}$ behaviour for $\nu \ll \nu_0$ produced by
the viscoelastic damping (see Eq.\,\ref{eq5}). However, the
measurement is still too noisy in order to study accurately
violations of the FDT during aging.

\begin{figure}
    \begin{center}
    \includegraphics[width=8cm, angle=0]{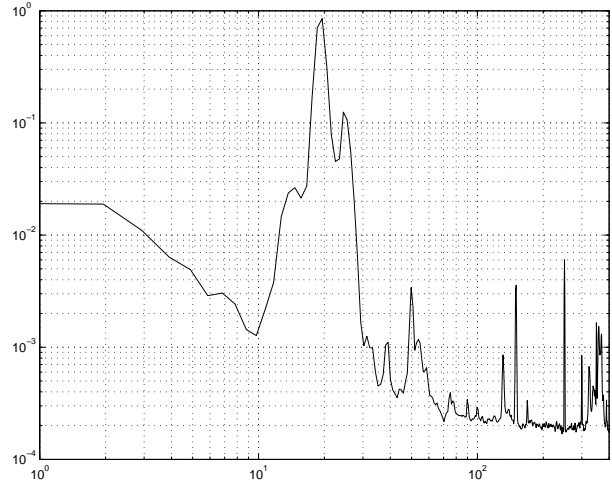}
    \caption{Square root of the spectral density of $x_{\mathrm{4Q}}$ vs frequency (Hz).
    This signal is related to the environmental noise.}
    \end{center}
    \label{fig4}
\end{figure}

\begin{figure}
    \begin{center}
    \includegraphics[width=8cm, angle=0]{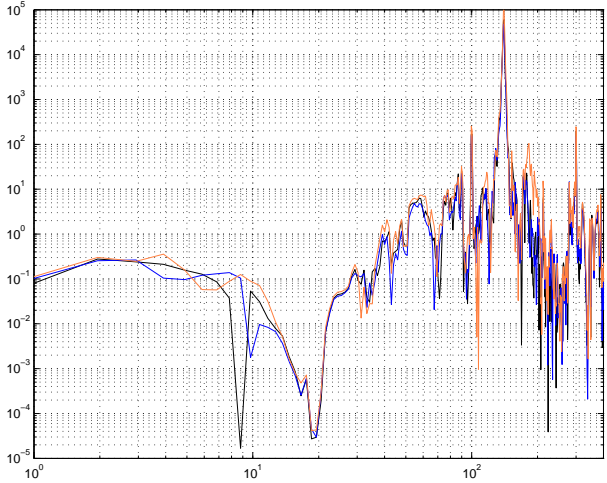}
    \caption{Example of three $\vert \hat{R}_{\mathrm{d}} \vert$ measured at different
    time intervals of length $ {T} = 1 \textrm{ min}$ separated by $2 \textrm{ min}$.
    Notice the large variation between the three curves taken at different times.}
    \end{center}
    \label{fig5}
\end{figure}

To improve our signal-to-noise ratio we have applied our
denoising technique described in Sec.\,\ref{sec2}. As already
mentioned, the total cantilever displacement reads
$x_c = x_T + x_{\mathrm{acc}}$. To get $x_T$ we have to estimate
$x_{\mathrm{acc}}$. The residual acceleration of the table where the
interferometer is installed is about, $10^{-8} \textrm{ m} \, \textrm{s}^{-2}$
at 1 Hz and $10^{-7} \textrm{ m} \, \textrm{s}^{-2}$ at 100 Hz.
This is too small to be detected by standard accelerometers, so we used
a different method. We built another cantilever made by harmonic steel
(cantilever C2) which is installed very close to the Polycarbonate
cantilever (cantilever C1). The parameters $\{ a, b, l, m \}$ of C2 are
chosen to optimize the sensitivity to mechanical vibrations and reduce its
sensitivity to thermal noise (see Eq.\,\ref{SNR}). A heavy mass and a heavy
mirror, that give the main contribution to $m$ for C2, are fixed on the steel
cantilever free end. The cantilever C2 is damped by the viscosity
of the air. A laser beam is reflected by the mirror glued on C2 and sent to a
four quadrant position sensitive photodiode (4Q), which is used to detect the
vibrations of the steel cantilever. The sensitivity is much
smaller than that of the Nomarski interferometer, but enough for
reducing the noise. Specifically, C2 is 20 mm long, 10 mm wide,
0.125 mm thick, it has a total mass of 1.3 g approximately and a resonance
frequency around 20 Hz. The maximum sensitivity external acceleration of this
setup, which is limited by the 4Q detector, is about $10^{-7} \textrm{ m} \,
\textrm{s}^{-2}$ in the frequency range of our interest. The output signal
$x_{\mathrm{4Q}}$ of the four quadrant detector and its Fourier transform
$\hat{x}_{\mathrm{4Q}}$ are mainly proportional to the response of C2 to the
external mechanical noise. Indeed, as we have already mentioned, thermal
fluctuations of C2 are negligible. An example of the square root of the spectral
density of the 4Q signal $x_{\mathrm{4Q}}$ is plotted in Fig.\,\ref{fig4},
which is related, via the response of C2, to the spectrum of
the residual acceleration of the optical table. The polycarbonate cantilever
and the steel cantilever, which are mounted very close on the same optical
table, are perturbed by the same environmental noise sources. As theses
sources may change of nature and of position, the responses of C1 and C2 to
these external perturbations may change too. That is the reason why the
denoising technique proposed in Sec.\,\ref{sec2} can be very useful, because
no hypothesis is needed on the response of the devices
to the external noise. Referring to Sec.\,\ref{sec2}, one has to make the
following substitutions: $x \rightarrow x_c$, $x_{\mathrm{true}}\rightarrow x_T$
and $x_{\mathrm{det}} \rightarrow x_{\mathrm{4Q}}$, whence
\begin{eqnarray}
    \langle {\vert \hat{x}_T \vert}^2 \rangle & = & \langle {\vert \hat{x}_c \vert}^2 \rangle - {\vert \hat{R}_{\mathrm{d}}\vert}^2 \langle
    {\vert \hat{x}_{\mathrm{4Q}} \vert}^2 \rangle, \\
    \textrm{with } \hat{R}_{\mathrm{d}} & = & \frac{\langle \hat{x} \,
    \hat{x}_{\mathrm{4Q}}^{*}\rangle}{\langle    {\vert \hat{x}_{\mathrm{4Q}} \vert}^2 \rangle}, \label{denois4Q}
\end{eqnarray}

\noindent where the average $\langle \cdot \rangle$ is computed in
our experiment on a time interval $ {T} = 1 \textrm{ min}$,
because $\hat{R}_{\mathrm{d}}$ evolves on a time scale of a few
minutes. This is shown in Fig.\,\ref{fig5} where we plot
$\hat{R}_{\mathrm{d}}$, measured in three different time intervals
separated by a few minutes. We see that the large variability of
this response will make any  a priori hypothesis useless. Using
these data, we apply the denoising technique and we compute
$\langle {\vert \hat{x}_T \vert}^2 \rangle$ for each time interval
of length $ {T}$. Finally, we average the spectra obtained over
several time intervals.

%
%

\begin{figure}
    \begin{center}
    \includegraphics[width=8cm, angle=0]{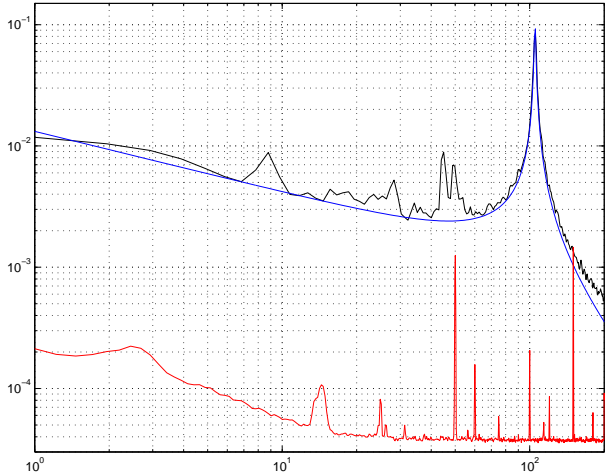} 
    \caption{Square root of the fluctuation spectral density with denoising
    ($\textrm{nm} / \sqrt{\mathrm{Hz}}$) vs frequency (Hz).}
    \label{fig6}
    \end{center}
\end{figure}

In Fig.\,\ref{fig6} we plot $\sqrt{ \langle \vert{ \hat{x}_T
\vert}^2 \rangle }$ as obtained after having applied the noise
reduction technique on twenty time intervals of length $ {T} = 1
\textrm{ min}$. By comparing this curve with Fig.\,\ref{fig3} we
see that all the peaks have been strongly reduced and that the
agreement with the FDT prediction is much better than in
Fig.\,\ref{fig3}. Notice that no improvement is observed if the
denoising technique is applied on a single time interval of $ {T}
= 20 \textrm{ min}$. We stress again that this effect is due to
the fact that response is changing as a function of time.

This example clearly shows that the denoising technique proposed
in Sec.\,\ref{sec2} can reduce the influence of environmental noise on a
measure if $R_{\mathrm{d}}$ is computed on short time intervals. The strong
noise reduction introduced by this technique allows us to study
the evolution of the FDT in an aging material. The accuracy is now
limited by the 4Q noise, but this can be strongly reduced by
replacing it with another Nomarski interferometer.

\section{Discussion and conclusions} \label{sec5}

In this article, we have proposed an original and simple denoising
technique, which allows one to reduce the influence of the
environmental noise on a measure. As already mentioned, this
denoising technique mixes several features of the standard
feedback systems and those of the Wiener filtering. The example
presented in Sec.\,\ref{sec3} clearly shows that this technique can be very
effective in suppressing spurious peaks on the spectra. The
example of Sec.\,\ref{sec3} is not exhaustive. Indeed, the same technique can
be used to reduce pick-up effects in electrical measurements or
eventually in very precise AFM measurements.

As a general conclusions we can say that this technique is simple
and can be implemented rather easily. The only requirement is to
have a reliable measurement of the environmental noise. Of course,
it can be strongly improved by a multidirectional measurement of
the noise.

\vspace{0.5cm}
\noindent \textbf{Acknowledgements}
\vspace{0.5cm}

The authors thank P. Abry, L. Bellon and I. Rabbiosi for useful discussions, and acknowledge
M. Moulin, F. Ropars and F. Vittoz for technical support.


\begin{thebibliography}{99}

\bibitem{lindgren}  ETS-LINDGREN Documentation, \textit{RF Shielded Enclosures,
                    Modular Shielding System Series} $81^{\textrm{TM}}$ (2002)

\bibitem{papoulis}  A. Papoulis, \textit{Signal Analysis}, 4th printing, McGraw-Hill
                   (1988)

\bibitem{max}       J. Max, \textit{M\'ethodes et Techniques de Traitement du Signal et
                    Applications aux Mesures Physiques, Tome 2, Appareillages, M\'ethodes Nouvelles,
                    Exemples d'Applications}, 4th edition, Masson (1987)

\bibitem{lifermann} J. Lifermann, \textit{Les Principes du Traitement Statistique du Signal,
                    Tome 1, Les M\'ethodes Classiques}, Masson (1981)

\bibitem{kalman}    R.E. Kalman, \textit{A new approach to linear filtering and
                    prediction problems}, Transactions of the ASME,                    Journal of Basic Engineering \textbf{82 D}, 35-45 (1960)

\bibitem{kalmanb}   R.E. Kalman, R.S. Bucy, \textit{New results in linear filtering and
                    prediction theory}, Transactions of the ASME, Journal of
                    Basic Engineering \textbf{38 D}, 95-108 (1961)

\bibitem{israeloff} T.S. Grigera, N.E. Israeloff, \textit{Observation of fluctuation-dissipation-theorem violations
                    in a structural glass}, Phys. Rev. Lett. \textbf{83}, 5038-5041 (1999)



\bibitem{be3}       L. Bellon, S. Ciliberto, \textit{Experimental study of the fluctuation dissipation during an
                    aging process}, Physica D \textbf{168}-\textbf{169}, 325-335 (2002)


\bibitem{herisson}  D. H\'erisson, M. Ocio, \textit{Fluctuation-dissipation ratio of a spin glass in the aging regime},
                    Phys. Rev. Lett. \textbf{88}, 257702-1 - 257202-4 (2002)

\bibitem{be5}       L. Bellon, L. Buisson, S. Ciliberto, F. Vittoz, \textit{Zero applied stress rheometer}, Rev. Sci.
                    Instrum. \textbf{73} (\textbf{9}), 3286-3290 (2002)

\bibitem{lionel}    L. Buisson, L. Bellon, S. Ciliberto, \textit{Intermittency in ageing}, J. Phys.: Condens. Matter
                    \textbf{15}, S1163-S1179 (2003)

\bibitem{nr}        W.H. Press, S.A. Teukolsky, W. T. Vetterling, B.P. Flannery, \textit{Numerical Recipes in C.
                    The Art of Scientific Computing, Second Edition}, Cambridge University Press (1992)

\bibitem{cugliandolo}   L.F. Cugliandolo, J. Kurchan, L. Peliti, \textit{Energy flow, partial equilibration, and effective
                        temperatures in systems with slow dynamics}, Phys. Rev. E \textbf{55} (\textbf{4}), 3898-3914 (1997)






\bibitem{sollich}   S. Fielding, P. Sollich, \textit{Observable dependence of the fluctuation dissipation relation and effective
                    temperature}, Phys. Rev. Lett. \textbf{88} (\textbf{5}), 050603-1 - 050603-4 (2002)

\bibitem{bib1}      L.D. Landau, E.M. Lifshitz, \textit{Theory of Elasticity},
                    3rd edition, Butterworth-Heinemann (1986)

\bibitem{bib2}      L.D. Landau, E.M. Lifshitz, \textit{Statistical Physics}, Part 1,
                    3rd edition, Butterworth-Heinemann (1980)

\bibitem{bib3}      N.G. McGrum, B.E. Read, G. Williams, \textit{Anelastic and
                    Dielectric Effects in Polymeric Solids}, Wiley (1967)

\bibitem{bib4}      G. Nomarski, \textit{Microinterf\'erom\`etre \`a ondes polaris\'ees}, J. Phys. Radium \textbf{16}, 9S-16S (1954)

\bibitem{bib5}      M. Fran\c con, S. Mallick, \textit{Polarization
                    Interferometers}, Wiley (1971)

\bibitem{bib6}      L. Bellon, S. Ciliberto, H. Boubaker, L. Guyon, \textit{Differential interferometry with a complex contrast},
                    Optics Communications \textbf{207}, 49-56 (2002)

\bibitem{bib7}      G. Ballardin et al, \textit{Measurement of the VIRGO superattenuator performance for seismic noise
                    suppression}, Rev. Sci. Instrum. \textbf{72} (\textbf{9}), 3643-3652 (2001)

\bibitem{bib8}      G. Losurdo et al, \textit{Inertial control of the mirror suspensions of the VIRGO interferometer
                    for gravitational wave detection}, Rev. Sci. Instrum. \textbf{72} (\textbf{9}), 3653-3661 (2001)

\bibitem{bib9}      E. Coccia, V. Fafone, \textit{Noise attenuators for gravitational wave experiments}, Nucl. Instr. and
                    Meth. in Phys. Res. A \textbf{366}, 395-402 (1995)

\bibitem{bib10}     E. Puppin, V. Fratello, \textit{Vibration isolation with magnet springs}, Rev. Sci. Instrum.
                    \textbf{73} (\textbf{11}), 4034-4036 (2002)

\end{thebibliography}
\end{document}